\documentclass[12pt,aasms5]{aastex}

\def\ltsima{$\;\buildrel < \over \sim \;$}
\def\simlt{\lower.5ex \hbox{\ltsima}}
\def\gtsima{$\;\buildrel > \over \sim \;$}
\def\simgt{\lower.5ex \hbox{\gtsima}}
\def\hbar{\mathrel{\hbox{\rlap{\hbox{\raise3pt\hbox{$-$}}}\hbox{$h$}}}}
\begin{document}
\title{Photon Orbital Angular Momentum in Astrophysics}
\author{Martin Harwit} 
\affil{511 H St., SW, Washington, DC 20024; also Cornell University, harwit@verizon.net}

\begin{abstract}
Astronomical observations of the {\it orbital angular momentum of photons}, a property of
electromagnetic radiation that has come to the fore in recent years, have apparently never been
attempted.  Here, I show how measurements of this property of photons have a number of
astrophysical applications. 
\end{abstract}


\section{Introduction}

Photons are endowed with spin angular momentum $\pm\hbar$ along their direction of
propagation.  Beams of photons all carrying the same spin are circularly
polarized.   Less well known is that photons can also carry orbital angular momentum (OAM),
$\ell$, quantized in units of $\hbar$.  Curtis, Koss \& Grier (2002) have produced beams of
photons each with OAM as high as $\ell = 200 \hbar$,

Progress in laboratory studies of photon orbital angular momentum (POAM) has been rapid since
Allen, Beijersbergen, Spreeuw \& Woerdman (1992) first pointed out that laser --- and by
inference maser --- modes
with well-defined POAM can be readily produced.  The characteristics of this radiation are by
now reasonably well established (Allen, Padgett \& Babiker, 1999; Allen, 2002).

A new development within the last year has been the introduction of a straightforward technique
for measuring the OAM of individual photons (Leach et al, 2002).  Rather than measuring the
angular momentum of the photons directly, the new method sorts photons
according to their symmetry properties.  This should permit the introduction of measurements of
POAM into astronomy. 

In section 2 of this paper, I provide a short quantitative introduction to POAM, followed in
section 3 by a description of the astronomical instrumentation required to detect POAM and
several of its limitations.  Section 4 lists a range of  astrophysical observations
that could be undertaken.  Section 5 briefly summarizes the findings.

\section{Multipole Fields}

Spherical electromagnetic waves in free space, like all waves entailing divergence-free fields,
$\nabla \cdot {\bf E} = \nabla \cdot {\bf H} = 0$, can be completely described by superpositions
of
electric and magnetic multipole fields.  For an electric multipole the magnetic field is transverse
to the direction of propagation (TM mode), while for a magnetic multipole the electric field is
transverse (TE mode). The TE and TM modes are dual to each other, related through the
transforms  
\begin{equation}
{\rm E}^{(E)} \rightarrow - {\rm H}^{(M)}\quad {\rm and}\quad  {\rm H}^{(E)} \rightarrow
{\rm E}^{(M)}
\end{equation}
where the superscripts (E) and (M) respectively indicate TE and TM modes (Rose, 1955). The
two modes correspond to two orthogonal senses of polarization (Jackson, 1975, P 398).

Classically the angular momentum of an electromagnetic wave is given by the volume integral of
the
cross product of position {\bf r} measured from the
center of the multipole
and the Poynting vector {\bf S} at {\bf r} 
\begin{equation}
{\bf J} = \frac {1}{c^2}\int {\bf r}{\bf \wedge} {\bf S} dV= \frac{1}{4\pi c}\int{\bf r}{\bf
\wedge}({\bf
E}{\bf \wedge}{\bf H})dV\ .
\end{equation} 
The same expression holds in quantum electrodynamics, but the vector field strengths now
become operators acting on a state vector $\Psi$. ${\bf J}$ can give rise to two components
which may not always be clearly separable,  ${\bf J} = {\bf J_o} + {\bf J_s}$,
respectively the orbital and spin angular momenta.\footnote{For different views on this
separability and its dependence on gauge invariance, see Jauch \& Rohrlich (1955), P40;
Gottfried (1966), page 412; Allen, Padgett \& Babiker (1999), p 304.}  

Quantum mechanically, one writes
\begin{equation}
{\bf J} = -i\hbar{\bf r\wedge \nabla}\quad +\quad \hbar{\bf s}\ , 
\end{equation}
where {\bf s} is the spin matrix for a vector field (Franz, 1950; Rose, 1955). The two angular
momentum eigenvalue equations deriving from the angular momentum operator are 
\begin{equation}
J_z {\bf E} =  m\hbar{\bf E} \quad {\rm and}\quad J^2{\bf E} =  \ell(\ell + 1)\hbar^2{\bf E}\ .
\end{equation}

From (2) it is clear that only the radial components of ${\bf E}$ or ${\bf H}$ can contribute to
the net angular momentum of a wave.  For electric multipole radiation these
components take the form (Heitler, 1936, 1954)
\begin{equation}
H_r = 0\ ,\quad E_r =  \frac{{\cal A}_{\ell}^ m S(kr)}{r(kr)^{1/2}} e^{-i\omega
t)}P_{\ell}^m(\cos\theta)
e^{im\phi}
\end{equation}
where ${\cal A}_{\ell}^m$ reflects the amplitude of the multipole, $S(kr)$ is derived from
Bessel functions $J_{\ell + 1/2}$, and $P_{\ell}^m$ is the associated Legendre
polynomial. $\ell$ and $m$ are integers with $|m|\leq \ell$.  $\omega$ is the angular frequency
and $k$ the wave number.  The multipole axis lies along some
direction ${\bf \epsilon}_z$, to which the vector ${\bf r}$ is inclined at an angle $\theta$, and
$\phi$ is the
azimuthal angle about the $z$-axis.

Classically, a vibrational motion along the multipole axis radiates perpendicular to this axis,
while a rotation about the $z$-axis produces radiation with angular momentum directed along the
multipole axis. (Morette De Witt \& Jensen, 1953).  \footnote{For further
discussions, see Heitler (1954), Gottfried (1966), and Jackson (1975).  The contemporary
literature on POAM, concerned largely with laser optics, uses the symbol $\ell$ for the magnetic
quantum number $m$.  This is at variance with the earlier literature and customary usage in
physics. I have tried to avoid confusion by adhering to the notation of the three cited books.}  Quantum mechanically, the $z$-component of the first term in (3) can be written as $J_z =
-i\hbar\delta/\delta\phi$ (Jackson,
1975, P 743), showing that the term $e^{im\phi}$ in equation (5) gives rise to an angular
momentum component $m$ about the $z$-axis.  In contrast, the spin component of the angular
momentum in (3)
is Lorentz invariant, always takes on the value $\pm\hbar$, and is directed along the axis of
propagation.  

The propagating electromagnetic wave consists of $m$ intertwined helical wave fronts, and  
$m$ is called the {\it winding number} or {\it topological
charge}.  All phases $\phi$ appear along the beam axis, $r = 0$, and the resulting destructive
interference leads to zero intensity there. Constructive interference occurs at some radius $r_m$
off the beam axis, so that light brought to a focus forms a ring of radius  
\begin{equation}
r_m = \frac{a\lambda f}{\pi \rho}\left(1 + \frac{m}{m_0}\right)\ ,
\end{equation} 
and width comparable to the
wavelength $\lambda$. Here $f$ the focal length, $\rho$ the radius of the optical train's
effective aperture, and the values $a\sim 2.585$ and $m_0 \sim 9.80$ are experimentally
determined (Curtis \& Grier, 2003).  In the limit of low values of $m$ the ring has dimensions
small compared to the Airy disk.  
  
Barnett and Allen (1994) examined the general
relationship between energy and angular momentum along the direction of propagation for
electromagnetic radiation and obtained the expression
\begin{equation}
\frac{J_z}{{\cal E}} = \frac{(m + \sigma)}{\omega} +\frac{\sigma}{\omega}[g(k)].
\end{equation}
$g(k)<< 1$ reflects the spectral and spatial distribution of
the radiation, and tends to be negligibly small. 

From (7) we see that though the spin and orbital angular momenta cannot be cleanly separated, it
is always possible to measure the orbital angular momentum by passing a beam through a linear
polarizer, which sets $\sigma = 0$, and leaves the orbital angular momentum intact.  The ratio
$J_z/{\cal E}$ then is $m /\omega$.

\section{Astronomical Instrumentation to Measure POAM}

\subsection{Dove-Prism Mach-Zehnder Interferometers}

With a relatively simple experiment, He et al (1995) were the first to show that POAM can be
transferred to small particles.  Working with a linearly polarized helium-neon laser
beam that could be switched between $m$ values $+3$ and $-3$ they set finely divided CuO
grains
suspended in water into clockwise or counterclockwise rotation, at will. 

Experiments by O'Neil et al. (2002) and Curtis \& Grier (2003) have clearly shown that the angular momentum absorbed by such small particles is orbital angular momentum, rather than spin.  They trapped microscopic particles in the highly focused annular image of radius $r_m$ produced by a laser beam with POAM $m=40$.  The particles then circled the optical axis along this annulus.

Until recently, however, there was no straightforward method for measuring the OAM of a single
photon with unknown $m$.  The provision of such a method now opens up for astronomy a
technique that should prove itself valuable.

The method has been described by Leach et al. (2002). Their apparatus was designed to deal with
laser-generated modes, but the procedure is general though instrumental details will differ for
different wavelength ranges and applications. The method does not directly
measure POAM.  Instead, it identifies the symmetry properties of a beam of electromagnetic
radiation subjected to a sequence of rotations about its axis of propagation.  This is achieved by
sending light through a cascade of Mach-Zehnder
interferometers with Dove prisms in each arm (Fig. 1).  At each stage the beams in the two
interferometer
arms are rotated with respect to one another through an angle $\alpha$, where $\alpha/2$ is the
relative rotation of the Dove prisms about the optical axis in each beam.  The first interferometer
stage has $\alpha/2 = \pi/2$ and sorts photons with even values of
orbital angular momentum $m$ into one exit port and those with odd values of $m$ into the
other port. The photons with odd values of $m$ are then sent through a hologram (Fig. 2(b)) that
increases the POAM $m$ carried by each photon to a
value $m +1$, thus endowing all the photons with even values of $m$.  

Each of the beams emerging from the two ports of this first stage is then sent through a second
Mach-Zehnder stage of its own, in which the two Dove prisms are rotated by an angle $\alpha/2
= \pi/4$ relative to each other.  These two stages, respectively dedicated to what originally were
odd and even modes $m$, now separate modes with $m = 4n$, where $n$ is an integer, from
those with $m = 4n+2$.  This process is continued in successive Mach-Zehnder stages until
photons with all desired values of $m$ have been sieved out.  Leach et al., (2002) have
demonstrated that the method provides clean separation for individual photons with m = 0, 1, 2,
and 3 when passed through a two-stage apparatus.  In principle, the Dove prisms could be
replaced by equivalent all-reflective elements for use over wide wavelength ranges.  

The winding number $m$ is invariant under a Lorentz
transformation.  This makes it a robust indicator of the orbital angular momentum.   

\subsection{Astronomical Limitations}

The theory discussed thus far can be applied to the detection of radiation from individual atoms,
molecules or lasers.  However, use of the technique of Leach et al. (2002) requires rotating a
beam with multipole characteristics ($\ell, m$) around the multipole symmetry axis $z$.  The
angular distribution of the radiation about this axis is (Blatt \& Weisskopf, 1952, page 594)  
\begin{equation}
\Omega_{\ell}^m(\theta,\phi)= \frac{1}{2}\left[ 1 - \frac{m(m+1)}{\ell(\ell +
1)}\right]|Y_{\ell}^{m + 1}|^2 + \frac{1}{2}\left[ 1 - \frac{m(m-1)}{\ell(\ell +
1)}\right]|Y_{\ell}^{m - 1}|^2 + \frac{m^2}{\ell (\ell + 1)}|Y_{\ell}^m|^2
\end{equation}
where the normalized spherical harmonics are 
\begin{equation}
Y_{\ell}^m(\theta,\phi) = \left [\frac{2\ell + 1}{4\pi} \frac{(\ell -|m|)!}{(\ell +|m|)!}\right
]^{1/2}P_{\ell}^m(\cos\theta)e^{im\phi}\ .
\end{equation}
Equation (8) holds for pure modes.  A superposition of modes can lead to interference effects
affecting the angular distribution.  

The associated Legendre polynomials $P_{\ell}^m$ have a
deep null along the multipole axis
for all quantum numbers $m \neq 0$, irrespective of $\ell$.  This means that astronomical
radiation incident on a telescope is extremely weak near the multipole axis unless $m = 0$ or,
alternatively, the beam is highly collimated, as for masers, and a helical structure is imposed on
the beam through an azimuthal phase shift.  This limitation appears to be universal, and extends
to other means for determining POAM.

\section{Astrophysical Applications}

Despite these limitations a number of astrophysical applications emerge. 

\subsection{Masers as Probes of Inhomogeneities}

Observations by Bignall et al. (2003) dramatically illustrate the existence of large density
inhomogeneities in the interstellar medium on small scales.   They observed radio flux density
changes of up to 40\%, over a period as short as 45 minutes, from the quasar PKS 1257-326. 
Interstellar and circumstellar masers similarly tend to be associated with 
shocked
domains.  To reach Earth, the radiation traverses regions that may well have discontinuities
impressing OAM on transmitted electromagnetic waves.  These effects can be
sizeable because the refractive index $n$ of the interstellar medium is substantial.  The group
velocity
of the wave is $c/n = c[1 + \omega_p^2/\omega^2]^{-1/2}$, where $\omega$ is the angular
frequency
of the wave and  $\omega_p \sim 5.6\times 10^4 n_e^{1/2}$ rad s$^{-1}$ is the plasma
frequency.  For a cosmic-ray-induced ionization fraction $n_e/n_H\sim 10^{-6}$, a
delay of one wavelength is reached over a distance of 
\begin{equation}
D\sim 10^{12} \left(\frac{10^{-5}}{n_e/n_H}\right)\left(\frac{10^5\ {\rm
cm}^{-3}}{n_H}\right)\left(
\frac{20\ {\rm cm}}{\lambda}\right)\ {\rm cm}\ ,
\end{equation}
which is small compared to the dimensions of the turbulent region around an evolved star, where
masers are typically found at radial distances $10^{16}$ to $10^{17}$ cm.  A turbulent screen
with
significant density spikes, through and around which the maser beam has to pass, is therefore
likely to induce POAM.

To visualize the production of POAM by a maser beam passing through an inhomogeneous
medium, one can envision the beam illuminating a spiral phase plate (Fig. 2(a)).  The top surface
of
the plate is displaced by a height $s$ after a full azimuthal rotation $\phi = 2\pi$. At a radial
distance $r$ from the optical axis the local azimuthal slope of this surface is $\theta = s/2\pi r$.
On emerging from the phase plate a ray passing through $r$ is deflected by an angle $\Psi$,
where Snell's law for small angles gives $(\Psi + \theta) \sim n\theta$, and $n$ is the refractive
index of the plate.  It is easy to see that $\Psi \sim (n-1)\theta = 
(n-1)s/2\pi r$.  

Before entering the spiral phase plate, a photon's linear momentum is $h/\lambda$.  On exiting
the phase plate, the component of the photon's linear momentum in the azimuthal direction is
$p_{\phi} \sim h\Psi/\lambda$, and its
angular momentum about the optical axis is $J_z = rp_{\phi}\sim rh\Psi/\lambda\sim
(n-1)\hbar/s\lambda = m\hbar$.  Here the step height $s$ is chosen an integer multiple $m$ of
$\lambda/(n-1)$, so that $s = m \lambda/(n-1)$.  $J_z$ is independent of the radial distance $r$
at which radiation passes through the phase plate.   
  
A turbulent medium with discontinuities can be envisaged as a screen of such spiral phase plates. 
The analysis of spatial discontinuities may then entail tracking changes in the observed winding
number for individual circumstellar masers as the turbulent supersonic outflow from the parent
star progresses.  

\subsection{Luminous Point Sources}

Radiation emitted by luminous pulsars and quasars may also encounter density discontinuities in
traversing the immediate surroundings of these respective sources (cf. Zavala \& Taylor, 2003).
These discontinuities again can impose a twist on the radiation, similar to that produced by a
spiral phase plate or a holographic phase plate.  Here, as in the example of the maser cited above,
the axis of propagation reaching the telescope is defined by the line of sight from the source to
the telescope, and the discontinuity inducing the POAM lies along this line of sight.  The
effective multipole axis is, therefore, collinear with the axis of propagation, and measurement of
the POAM is feasible.

An earmark of discontinuities in a plasma is that $J_z \propto (n -1)/\lambda
\propto\omega_p^2/\omega$ for a phase plate with step height $s$.  Since ${\cal E}\propto
\omega$ this means that $J_z/{\cal E}\propto \omega^{-2}$ and by (7) $m\propto \omega^{-1}
\propto \lambda$.

\subsection{SETI}

A number of investigators have recently turned to visual wavelengths in their Search for
Extraterrestrial Intelligence, SETI.  This comes at a time when researchers in optical
communication have discovered significant advantages that radiation with high values of
POAM might have for communication and quantum computing.  The ability to encode a single
photon with $\log_2N$ bits of information, by endowing it with a POAM of $N\hbar$ in place of
the conventional single bit of information granted by photon spin, carries great promise while
also providing possibilities for entanglement (Vaziri, Weihs \& Zeilinger, 2002).  For SETI an
additional advantage would be the absence of naturally occurring optical photons with high
POAM. Artificially generated photons would thus be more readily culled out from naturally
occurring diffuse optical radiation in space.

In order to measure the POAM of a SETI transmission, the observer will again need to gather
radiation surrounding the multipole axis of the propagating beam. This may be achieved with 
arrays of telescopes both at the transmitting and receiving ends, in order to keep the beam
sufficiently narrow.  For a separation $D$ between transmitters and receivers operating at a
wavelength $\lambda$ the array baseline $d$ would have to be $d \sim 170 (\lambda D)^{1/2}$
km, where $\lambda$ is measured in microns and $D$ in parsecs.  For planets the size of the
Earth, this should permit access over distances of order a hundred parsecs.  At the receiving
end, the incoming radiation would be directed to a central station where it could be
rotated around its multipole axis, as in the method of Leach et al. (2002).  O'Neil et al. (2002) have studied the case where the optical axis of the beam is not precisely centered on the receiving optics.  The orbital angular momentum density is affected, but the winding number should remain unaltered.

\subsection{Transfer of OAM by Kerr Black Holes} 

Teukolsky (1972) first looked in detail at the interaction of electromagnetic radiation with Kerr
black holes.  Mashhoon (1973) soon thereafter pointed out that electromagnetic radiation
scattered off black holes would absorb some of the hole's angular momentum.  More recently
Falcke, Melia \& Agol (2000) have discussed observations to detect the shape of a dark shadow
in the immediate vicinity of a rapidly rotating black hole.  To date, however, the transfer of
angular momentum to POAM has not been discussed.  

Angular momentum transfer and gravitational lensing of electromagnetic radiation both are
wavelength independent.  Angular momentum transfer from a Kerr black hole might be expected
to induce a helical form on an
incident wave, just as a phase plate does, except that the angle $\Psi$ in Fig. 1 would be
independent of wavelength. More specifically, radiation from a distant unresolved source
incident on a Kerr black hole along its axis of
symmetry and lensed by the hole should leave
the right side of equation (7) independent of wavelength and exhibit orbital angular momentum
$m$ proportional to $\omega$ or, equivalently, $\lambda^{-1}$.  A full theoretical
investigation of such effects would be of interest as a guide to searches in MACHO surveys.

\subsection{Blackbody Radiation and the Cosmic Microwave Background}  

A question that arises is how the existence of an additional set of well defined OAM quantum
states could be compatible with the conventional partition function encountered in blackbody
radiation, $Z(\nu)d\nu = (8\pi \nu^2  d\nu/h^3)$, for radiation frequency $\nu$
in interval $d\nu$ and unit volume. 

We may consider isolating a narrow bandwidth of radiation emanating from a blackbody source,  
selecting a fine pencil of this radiation and passing this through a linear polarizer, before
permitting it to enter the cascade of Dove-Prism Mach-Zehnder stages introduced by Leach et al.
(2002).  Knowing the width of the pencil beam, the bandwidth and the polarization, permits
calculation of $Z(\nu)d\nu$.  By providing additional information about the quantum number
$m$, the cascade of Leach et al. (2002) would then appear to endow the
radiation with more degrees of freedom than the partition function permits, in apparent violation
of Bose-Einstein statistics and the Heisenberg uncertainty principle.

The error in this conclusion is that measurement of the photon orbital angular momentum
introduces a calculable
uncertainty in the direction of the Poynting vector, which corkscrews around the axis of
propagation and thus introduces an uncertainty in the lateral momentum. As the experiment of
Curtis \& Grier illustrates, radiation no longer focuses onto a point, but rather onto a ring.  In the
limit of high $m$ equation (6) makes clear that the circumference of the ring, i.e. the uncertainty
in the
lateral momentum increases linearly with $m$.  Whatever information we can gain about an
individual photon's orbital angular momentum corresponds to an identical loss of information
about its direction of propagation.

Blackbody radiation is a superposition of multipole fields of many orders, $\ell$, and we should
expect this to be true of the cosmic microwave background radiation, CMBR. \footnote{It is
important to note that we are here dealing with the multipole structure of the electric and
magnetic field components, rather than the more frequently encountered multipole expressions
for the CMBR surface brightness distribution.}  Our line of sight to the surface of last scattering
intersects this surface at right angles, so the radiation reaching us will have been emitted
perpendicular to this surface, i.e. along the multipole axis.  Equation (8), however, shows that
only radiation characterized by Legendre polynomials of form $P_{\ell}^0$ are
emitted along the multipole axis, and we should expect all the received radiation to exhibit
quantum number $m = 0$, corresponding to different values of $\ell$, mutually interfering and
indistinguishable.  This yields the correct partition function.  

If the multipole distribution of $m$ values, measured by the probability distribution $P(m)$ 
were to be found to significantly differ from $m = 0$ along some lines of sight to the CMBR, we
could gain information about discontinuities along those sight lines, as discussed in section 5.
Gravitational discontinuities would produce a dependence of
winding number $m \propto \omega$, distinguishing these from density discontinuities, for
example in shocked regions of intracluster plasma, which would exhibit $m\propto
\omega^{-1}$.   

\section{Conclusion}

Astronomical observations to detect photon orbital angular momentum appear to have never been
undertaken to date.  I have cited a few astrophysical observations that might be
attempted to gain new insight into different observational phenomena, and have listed several of
the problems facing the measurement of POAM. Further theoretical as well as observational
efforts will be required to clarify the unfamiliar interaction
of POAM with matter, gravitation and magnetic fields.

\section*{Acknowledgments}

I am greatly indebted to Prof. Les Allen for his incisive and helpful critique of this paper.  I also
thank Peter Nisenson, Harvey Moseley and Emil Wolf for their comments on an earlier version
of the manuscript.  Conversations with David Grier, Richard Lyon, Alex Kutyrev, John Mather, Harvey
Moseley,
J\"{u}rgen Stutzki and Edward Wollack were helpful and instructive.  My work in infrared
astronomy has been supported by contracts from NASA.

\section*{Figure Captions}

Fig. 1. First stage of a Dove-Prism Mach-Zehnder interferometer cascade for sorting photons
carrying different amounts of orbital angular momentum (after Leach et al., 2002).

Fig. 2.  Photon orbital angular momentum (POAM) produced by a spiral phase plate (a), and an
example of a holographic phase plate (b) (after Allen et al., 1999).  For details see the text.

\vfill\eject
\centerline{\bf References}
\vskip 0.1 true in 
{\hoffset 20pt
\parindent = -20pt

Abel, T., Bryan, G. L., \& Norman, M. L. 2002, Science, 295, 93

Allen, L., Beijersbergen, M. W., Spreeuw, R. J. C. \& Woerdman, J. P. 1992, Phys. Rev. A, 45,
8185 

Allen, L., \& Padgett, M. J. 200, Opt. Commun., 184, 67

Allen, L., Padgett, M. J., \& Babiker, M. 1999, Progress in Optics, XXXIX, 291

Allen, L. 2002, J. Opt. B: Quantum Semiclass. Opt., 4, S1 

Barnett, S. M., \& Allen, L. 1994, Opt. Comm., 110, 670 

Bignall H. E. et al. 2002, ApJ, 585, 653 

Blatt, J. M. \& Weisskopf, V. F. 1952, {\it Theoretical Nuclear Physics}, Wiley, New York

Curtis, J. E., \& Grier, D. G. 2003, PRL, 133901

Curtis, J. E., Koss, B.A., \& Grier, D. G. 2002, Opt. Commun., 207, 169 

Falcke, H. Melia, F. \& Agol, E. 2000, ApJ, 528, L13

Franz, W. 1950, Zeitschrift f. Physik, 126, 363

Fuller, T. M. \& Couchman, H. M. P. 2000, ApJ, 544, 6.

Gottfried, K. 1966, {\it Quantum Mechanics, Volume 1, Fundamentals}, Benjamin/Cummings 

He, H., Friese, M. E. J., Heckenberg, N. R., \& Rubinsztein-Dunlop, H. 1995, PRL, 75, 826 

Heitler, W. 1936, Proc. Camb. Phil. Soc., 32, 112

Heitler, W. 1954, {\it The Quantum Theory of Radiation} 3rd edition, Oxford, pages 401-4

Jackson, J. D. 1975, {\it Classical Electrodynamics}, John Wiley \& Sons, New York, Chapter 16

Jauch, J. M., \& Rohrlich, F. 1955, Addison-Wesley, Reading

Leach, J. Padgett, M. J., Barnett, S. M., Franke-Arnold, S. \& Courtial, J. 2002, PRL,
88, 257901

Mashoon, B. 1973, Phys. Rev., D 7, 2807

Morette De Witt, C. \& Jensen, J. H. D. 1953, Z. f. Naturforschung, 8a, 267

O'Neil, A. T., Mac Vicar, I., Allen, L. \& Padgett, M. J. 2002, PRL, 88, 053601

Rose, M. E. 1955, {\it Multipole Fields}, John Wiley \& Sons, New York

Teukolsky, S. A. 1972, PRL, 29, 1114

Vaziri, A., Weihs, G. \& Zeilinger, A. 2002, PRL, 89, 240401

Zavala, R. T. \& Taylor, G. B. 2003, ApJ, 589, 126

}

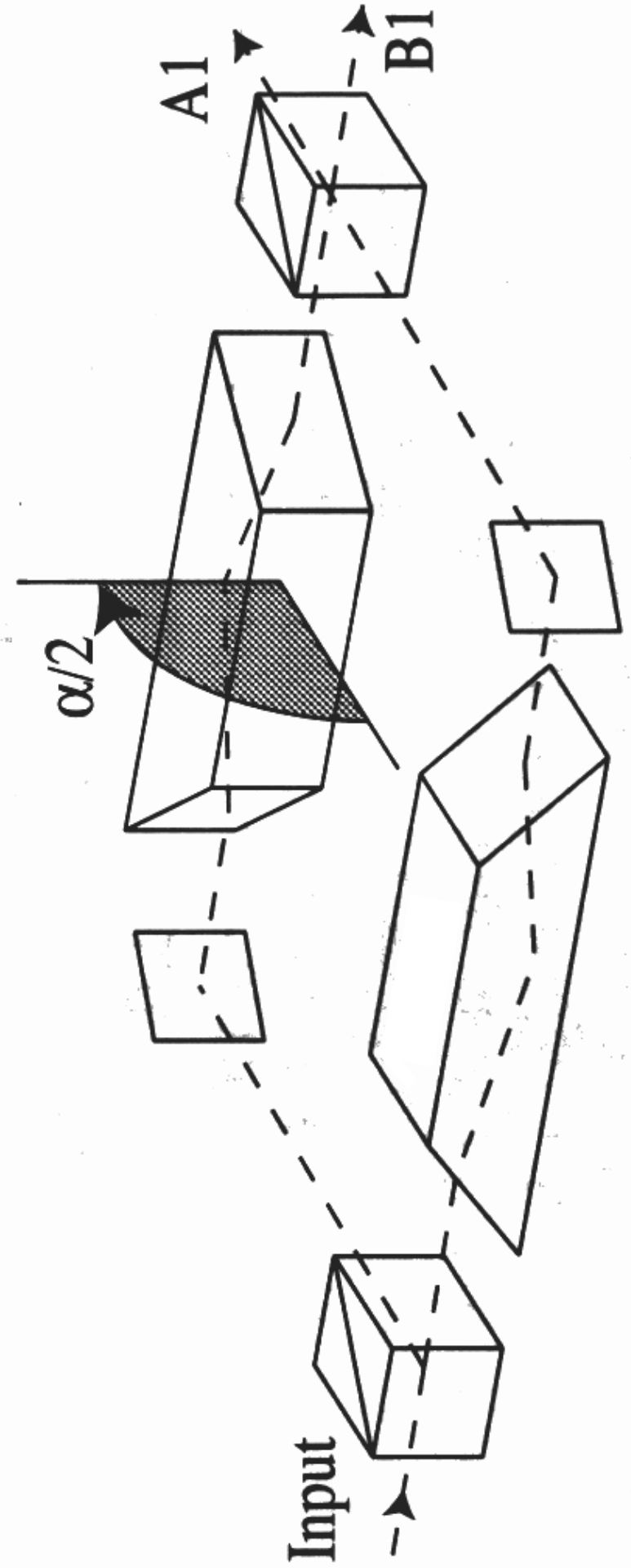
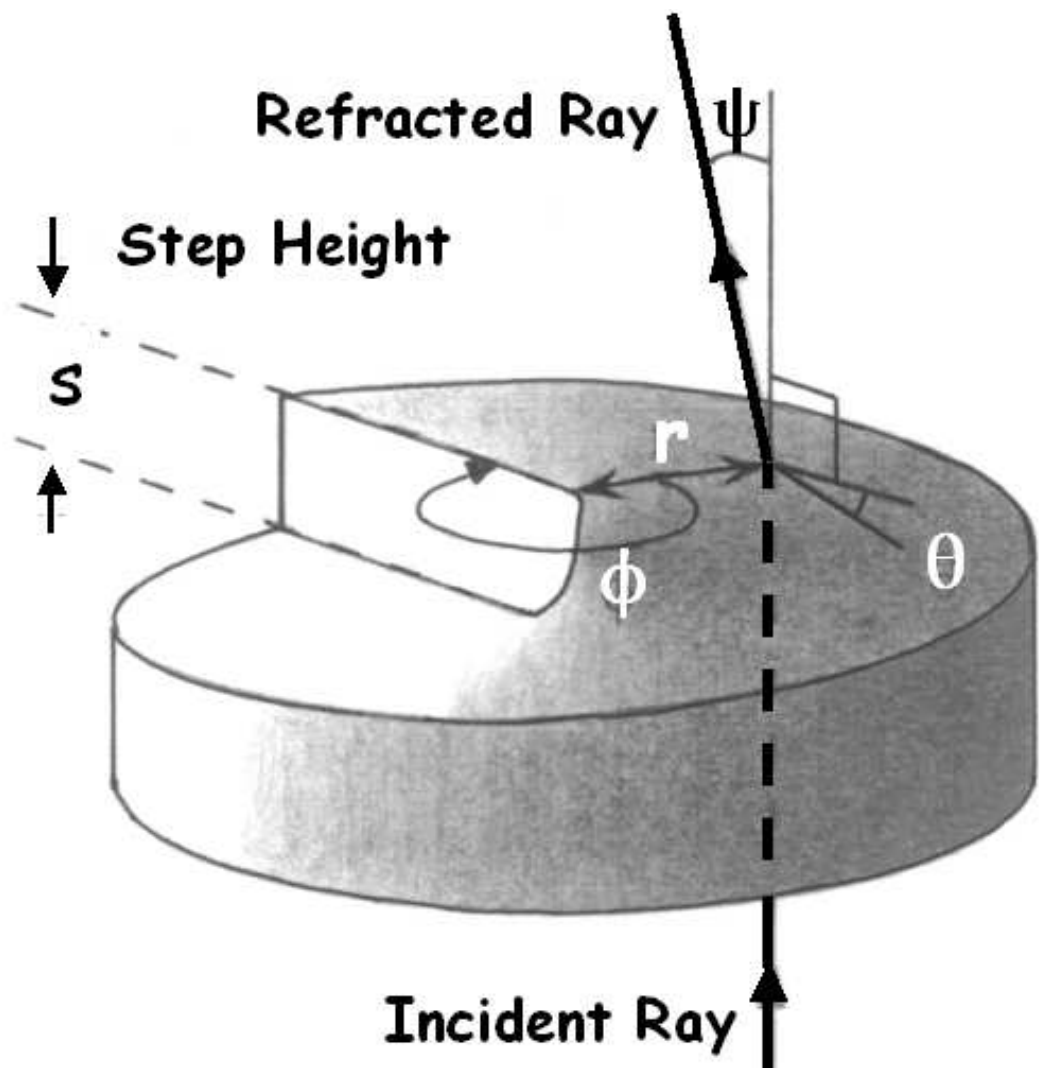
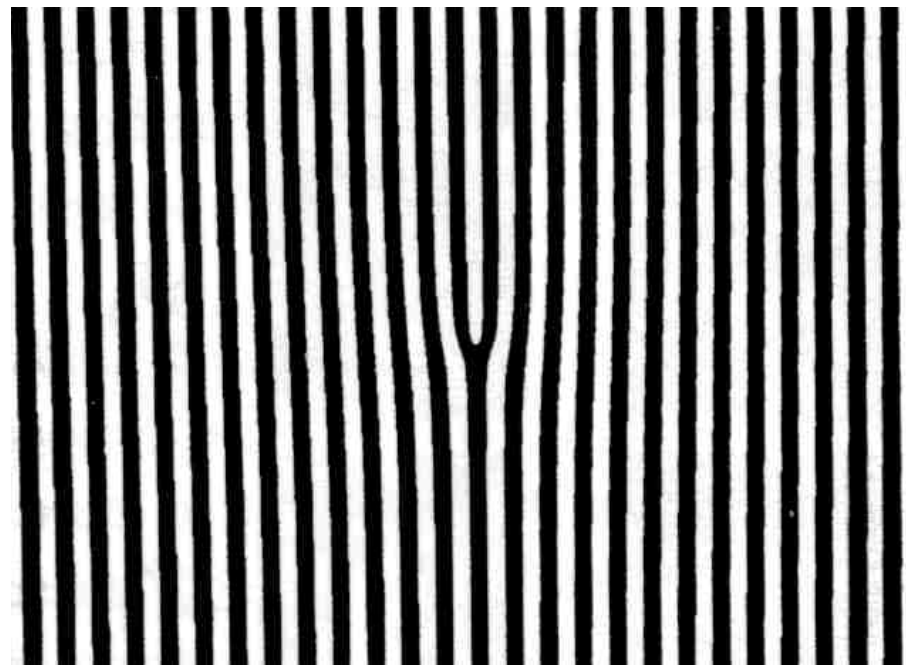
\end{document}